\documentclass[conference]{IEEEtran}

\usepackage[cmex10]{amsmath}
\usepackage{amssymb}

\usepackage{cases}
\usepackage{multirow}
\usepackage{empheq}

\usepackage{cite}
\usepackage{verbatim}

\ifCLASSINFOpdf
  \usepackage[pdftex]{graphicx}
\else
  \usepackage[dvips]{graphicx}
\fi
\usepackage[caption=false,font=footnotesize]{subfig}

\usepackage{color}
\usepackage{multirow}

\newtheorem{theorem}{Theorem}
\newtheorem{lemma}[theorem]{Lemma}

\newtheorem{corollary}[theorem]{Corollary}
\newtheorem{definition}[theorem]{Definition}

\newtheorem{remark}[theorem]{Remark}

\DeclareMathOperator{\rank}{rank}

\begin{document}

\sloppy

\title{Duality between Erasures and Defects}

%\author{
%\IEEEauthorblockN{Yongjune Kim and B. V. K. Vijaya Kumar}
%\IEEEauthorblockA{Data Storage Systems Center (DSSC)\\
%    Carnegie Mellon University\\
%    Pittsburgh, PA, USA\\
%    Email: yongjunekim@cmu.edu, kumar@ece.cmu.edu}
%\and
%\IEEEauthorblockN{Robert Mateescu}
%\IEEEauthorblockA{HGST Research\\
%San Jose, CA, USA\\
%Email: robert.mateescu@hgst.com}
%}

\author{\IEEEauthorblockN{Yongjune Kim and B. V. K. Vijaya Kumar}
\IEEEauthorblockA{Electrical and Computer Engineering, Carnegie Mellon University, Pittsburgh, PA, USA\\ Email: yongjunekim@cmu.edu, kumar@ece.cmu.edu}
}

%% To balance the two columns, you should reduce the text-height of
%% the last page using the following command:
%%%%%%%%%%%%%%%%%%%%%%%%%%%%%%%%%%%%%%%%%%%%%%%%%%%%%%%%%%%%%%%%%%%%%
%\addtolength{\textheight}{-9.35cm}
%%%%%%%%%%%%%%%%%%%%%%%%%%%%%%%%%%%%%%%%%%%%%%%%%%%%%%%%%%%%%%%%%%%%%
%% with an appropriate value. This command must be place on the second
%% last page, i.e., for a one-page abstract here, for a two-page
%% abstract right after the \maketitle command.

%% Create the title:
\maketitle

\begin{abstract}
	We investigate the duality of the binary erasure channel (BEC) and the binary defect channel (BDC). This duality holds for channel capacities, capacity achieving schemes, minimum distances, and upper bounds on the probability of failure to retrieve the original message. In addition, the relations between BEC, BDC, binary erasure quantization (BEQ), and write-once memory (WOM) are described. From these relations we claim that the capacity of the BDC can be achieved by Reed-Muller (RM) codes under maximum a posterior (MAP) decoding. Also, polar codes with a successive cancellation encoder achieve the capacity of the BDC. 
	
	Inspired by the duality between the BEC and the BDC, we introduce locally rewritable codes (LWC) for resistive memories, which are the counterparts of locally repairable codes (LRC) for distributed storage systems. The proposed LWC can improve endurance limit and power efficiency of resistive memories.  	
\end{abstract}

\section{Introduction}

	The binary erasure channel (BEC) is a very well known channel model, which was introduced by Elias~\cite{Elias1955}. Due to its simplicity, it has been a starting point to design new coding schemes and analyze the properties of codes. Moreover, the BEC is a very good model of for communications over the Internet and distributed storage systems. 
	
	In the BEC, the channel input $X \in \{0, 1\}$ is binary and the channel output $Y=\{0, 1, *\}$ is ternary. It is assumed that the decoder knows the locations of erased bits denoted by $*$. The capacity of the BEC with erasure probability $\alpha$ is given by \cite{Elias1955, Cover2006}
	\begin{equation}\label{eq:BEC_capacity}
		C_{\text{BEC}} = 1 - \alpha.
	\end{equation}
	
	Elias~\cite{Elias1955} showed that the maximum a posteriori (MAP) decoding of random codes can achieve $C_{\text{BEC}}$. In the BEC, MAP decoding of linear codes is equivalent to solving systems of linear equations whose complexity is $\mathcal{O}(n^3)$~\cite{Elias1955}. Subsequently, codes with lower encoding and decoding complexity were proposed~\cite{Luby2001,Shokrollahi2006,Arikan2009}. 
	
	The binary defect channel (BDC) also has a long history. The BDC was introduced to model computer memory such as erasable and programmable read only memories (EPROM) and random access memories (RAM) by Kuznetsov and Tsybakov~\cite{Kuznetsov1974}. Recently, the BDC has received renewed attention as a possible channel model for nonvolatile memories such as flash memories and resistive memories~\cite{Lastras-Montano2010, Jagmohan2010coding, Hwang2011iterative, Jacobvitz2013, Kim2013coding, Kim2014dirtyflash, Kim2015detrap}. 
	
	\begin{figure}[!t]
		\centering
		\includegraphics[width=0.33\textwidth]{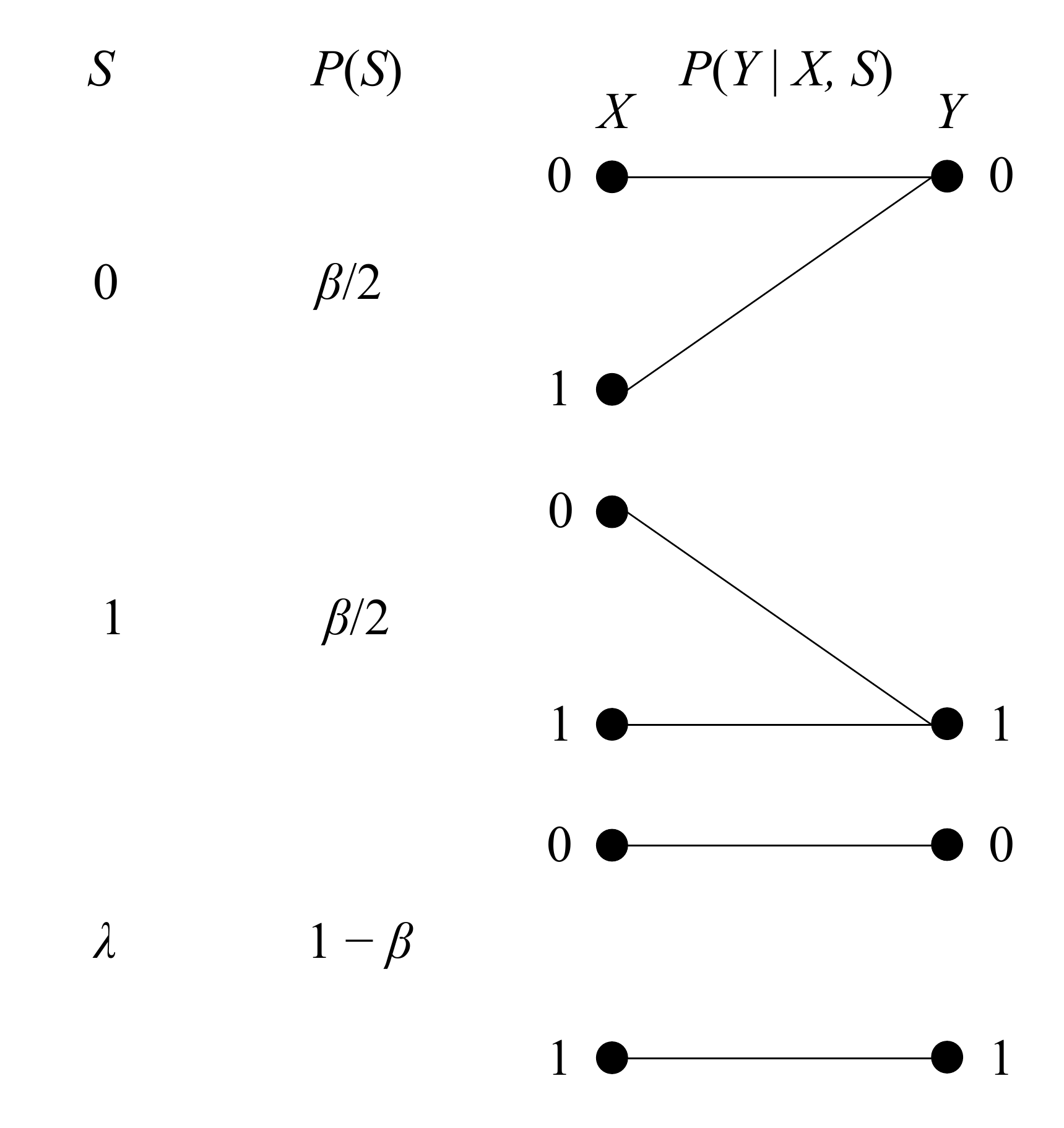}
		\caption{Binary defect channel (BDC).}
		\label{fig:BDC}
		\vspace{-3mm}
	\end{figure}
		
	As shown in Fig.~\ref{fig:BDC}, the BDC has a ternary channel state $S \in \{0, 1, \lambda\}$ whereas the channel input $X$ and the channel output $Y$ are binary. The state $S=0$ corresponds to a stuck-at 0 defect where the channel always outputs a 0 independent of its input value, the state $S=1$ corresponds to a stuck-at 1 defect that always outputs a 1, and the state $S = \lambda$ corresponds to a normal cell that outputs the same value as its input. The probabilities of these states are $\beta / 2$, $\beta / 2$ (assuming a symmetric defect probability), and $1 - \beta$, respectively~\cite{Heegard1983capacity, ElGamal2011}. 
	
	It is known that the capacity is $1 - \beta$ when both the encoder and the decoder know the channel state information (i.e., defect information). If the decoder is aware of the defect locations, then the defects can be regarded as erasures so that the capacity is $1 - \beta$~\cite{Heegard1983capacity, ElGamal2011}. On the other hand, Kuznetsov and Tsybakov assumed that the encoder knows the defect information (namely, the locations and stuck-at values of defects) and the decoder does not have any information of defects~\cite{Kuznetsov1974}. It was shown that the capacity is $1 - \beta$ even when only the encoder knows the defect information~\cite{Kuznetsov1974, Heegard1983capacity}. Thus, the capacity of the BDC is given by
	\begin{equation}\label{eq:BDC_capacity}
		C_{\text{BDC}} = 1 - \beta.
	\end{equation}
	
	The capacity of the BDC can be achieved by the \emph{binning scheme}~\cite{Heegard1983capacity, ElGamal2011} or the \emph{additive encoding}~\cite{Tsybakov1975additive, Dumer1990}. The objective of both coding schemes is to choose a codeword whose elements at the locations of defects match the stuck-at values of corresponding defects. %It is worth mentioning that binning scheme and additive encoding can be equivalent to solving systems of linear equations~\cite{Tsybakov1975additive, Zamir2002}. 
	
	We have studied the duality of erasures and defects and and our observations and results can be found in~\cite{Kim2014duality}. This duality can be observed in channel properties, capacities, capacity-achieving schemes, and their failure probability. In~\cite{Mahdavifar2015}, it was shown that we can construct capacity-achieving codes for the BDC based on state of the art codes which achieve $C_{\text{BEC}}$.  
	
	Recently, it was proved that Reed-Muller (RM) codes achieve $C_{\text{BEC}}$ under MAP~\cite{Kudekar2016}. Based on the duality of the BEC and the BDC, we show that RM codes can achieve $C_{\text{BDC}}$ with $\mathcal{O}(n^3)$ complexity. 
	
	Also, we extend this duality to the other models such as binary erasure quantization (BEQ) problems~\cite{Martinian2003iterative}, and write once memories (WOM)~\cite{Rivest1982wom}. We review the related literature and describe the relations between these models. From these relations, we can claim that $C_{\text{BDC}}$ can be achieved with $\mathcal{O}(n \log n)$ complexity which is better than the best known result in~\cite{Dumer1990}, i.e., $\mathcal{O}(n \log^2 n)$ complexity. 
		
	By taking advantage of this duality between the BEC and the BDC, we introduced locally rewritable codes (LWC)\footnote{LWC instead of LRC is used as the acronym of locally rewritable codes in order to distinguish them from locally repairable codes (LRC).} in~\cite{Kim2016lwc}. The LWC are the counterparts of locally repairable codes (LRC). The LRC is an important group of codes for distributed storage system~\cite{Huang2007pyramid, Gopalan2012} whose channel model is the BEC. On the other hand, the LWC is coding for resistive memories, which can be modeled by the BDC. 
	
	%The objective of LRC is to ensure fast repair by introducing \emph{repair locality} $r$. If an element of the LRC codeword is lost due to a single disk node failure among $n$ disk nodes, it can be repaired (i.e., reconstructed) by accessing at most $r$ other disk nodes~\cite{Gopalan2012, Tamo2014LRC}. Low repair locality enables fast repair in distributed storage systems. 
	
	%In the LWC, \emph{rewriting locality} $r^{\star}$ is defined as a counterpart of repair locality $r$. Suppose that there is only a single defect among $n$ memory cells. An element of the LWC codeword in the location of this defect can be updated by rewriting at most $r^{\star}$ other memory cells. By adopting LWC with low rewriting locality, we can improve endurance and power consumption of resistive memories. Based on the duality of erasures and defects, we will show that existing construction methods of LRC can be applied to construct LWC. 
	
	The rest of this paper is organized as follows. Section~\ref{sec:duality} discusses the duality between erasures and defects, which summarizes the results of \cite{Kim2014duality}. Also, the implications of this duality are investigated. In Section~\ref{sec:LWC}, we explain the LWC in~\cite{Kim2016lwc} and investigate the properties of LWC based on the duality of erasure and defects. Section~\ref{sec:conclusion} concludes the paper. 
	
\section{Duality between Erasures and Defects}\label{sec:duality}

\subsection{Notation}

We use parentheses to construct column vectors from comma separated lists. For a $n$-tuple column vector $\mathbf{a} \in \mathbb{F}_q^n$ (where $\mathbb{F}_q$ denotes the finite field with $q$ elements and $\mathbb{F}_q^n$ denotes the set of all $n$-tuple vectors over $\mathbb{F}_q$), we have
\begin{equation} \label{eq:vector}
	(a_1, \ldots, a_n) = \begin{bmatrix}
		a_1 \\ \vdots \\ a_n
	\end{bmatrix} = \left[a_1 \: \ldots \: a_n \right]^T
\end{equation}
where superscript $T$ denotes transpose. Note that $a_i$ represents the $i$-th element of $\mathbf{a}$. For a binary vector $\mathbf{a} \in \mathbb{F}_2^n$, $\overline{\mathbf{a}}$ denotes the bit-wise complement of $\mathbf{a}$. For example, the $n$-tuple all-ones vector $\mathbf{1}_n$ is equal to $\overline{\mathbf{0}}_n$ where $\mathbf{0}_n$ is the $n$-tuple all-zero vector. Also, $\mathbf{0}_{m, n}$ denotes the $m \times n$ all-zero matrix. 

In addition, $\| \mathbf{a} \|$ denotes the Hamming weight of $\mathbf{a}$ and $\text{supp}(\mathbf{a})$ denotes the support of $\mathbf{a}$. Also, we use the notation of $[i : j] = \{i, i+1, \ldots,j-1, j\}$ for $i < j$ and $[n] = [1:n]= \{1, \ldots, n \}$. Note that $\mathbf{a}_{[i:j]} = \left(a_i, \ldots, a_j\right)$ and $\mathbf{a}_{\setminus i} = (a_1, \ldots, a_{i-1}, a_{i+1}, \ldots, a_n)$.

\subsection{Binary Erasure Channel}

	For the BEC, the codeword most likely to have been transmitted is the one that agrees with all of received bits that have not been erased. If there is more than one such codeword, the decoding may lead to a failure. Thus, the following simple coding scheme was proposed in \cite{Elias1955}.

	\emph{Encoding:} A message (information) $\mathbf{m}\in \mathbb{F}_2^k$ is encoded to a corresponding codeword $\mathbf{c} \in \mathcal{C}$ where $ \mathcal{C} = \{ \mathbf{c} \in \mathbb{F}_2^n \mid \mathbf{c} = G \mathbf{m}, \mathbf{m} \in \mathbb{F}_2^k \}$ where $\mathcal{C}$ is a set of codewords and the generator matrix is $G \in \mathbb{F}_2^{n \times k}$ such that $\rank(G)=k$. Note that the code rate $R = \frac{k}{n}$.

	\emph{Decoding:} Let $g$ denote the decoding rule. If the channel output $\mathbf{y}$ is identical to one and only one codeword on the unerased bits, the decoding succeeds. If $\mathbf{y}$ matches completely with more than one codeword on the unerased bits, the decoder chooses one of them randomly~\cite{Elias1955}.

	We will define a random variable $D$ as follows.
	\begin{equation}
		D =
		\begin{cases}
			0,  & \mathbf{c} \ne \widehat{\mathbf{c}}\text{ (decoding failure)}; \\
			1,  & \mathbf{c} = \widehat{\mathbf{c}}\text{ (decoding success)}
		\end{cases}
	\end{equation}
	where $\widehat{\mathbf{c}}$ is the estimated codeword produced by the decoding rule of $g$.

	Elias showed that random codes of rates arbitrarily close to $C_{\text{BEC}}$ can be decoded with an exponentially small error probability using the MAP decoding~\cite{Elias1955, Shokrollahi2006,Richardson2008}. The MAP decoding rule of $g$ can be achieved by solving the following linear equations~\cite{Elias1955}:
	\begin{equation}\label{eq:BEC_decoder_LE}
		G^{\mathcal{V}} \widehat{\mathbf{m}} = \mathbf{y}^{\mathcal{V}}
	\end{equation}
	where $\widehat{\mathbf{m}}$ is the estimate of $\mathbf{m}$ and $\mathcal{V}=\left\{j_1,\cdots, j_v\right\}$ indicates the locations of the $v$ unerased bits. We use the notation of $\mathbf{y}^{\mathcal{V}}=\left(y_{j_1}, \cdots, y_{j_v}\right)$ and $G^{\mathcal{V}}=\left[ \mathbf{g}_{j_1}^T, \cdots, \mathbf{g}_{j_v}^T \right]^T$ where $\mathbf{g}_j$ is the $j$-th row of $G$. Note that $G^{\mathcal{V}} \in \mathbb{F}_2^{(n-e) \times k}$.
	
	The decoding rule $g$ can also be represented by the parity check matrix $H$ instead of the generator matrix $G$ as follows.
	\begin{equation}\label{eq:BEC_decoder_LE_H_1}
		H^T \mathbf{\widehat{c}} = \left(H^{\mathcal{E}} \right)^T \widehat{\mathbf{c}}^{\mathcal{E}} + \left(H^{\mathcal{V}} \right)^T \widehat{\mathbf{c}}^{\mathcal{V}} = \mathbf{0}
	\end{equation}
	where the parity check matrix $H$ is an $n \times (n-k)$ matrix such that $H^{T}G = \mathbf{0}$. Also, $\mathcal{E}=\left\{i_1,\cdots, i_e\right\}$ indicates the locations of the $e$ erased bits such that $\mathcal{E} \cup \mathcal{V} = [n]$ and $\mathcal{E} \cap \mathcal{V} = \emptyset$ (i.e., $n=e+v$). Note that $\widehat{\mathbf{c}}^{\mathcal{E}}=\left(\widehat{c}_{i_1}, \cdots, \widehat{c}_{i_e}\right)$, $\widehat{\mathbf{c}}^{\mathcal{V}}=\left(\widehat{c}_{j_1}, \cdots, \widehat{c}_{j_v}\right)$, $H^{\mathcal{E}}=\left[ \mathbf{h}_{i_1}^T, \cdots, \mathbf{h}_{i_e}^T \right]^T$ and $H^{\mathcal{V}}=\left[ \mathbf{h}_{j_1}^T, \cdots, \mathbf{h}_{j_v}^T \right]^T$ where $\mathbf{h}_i$ is the $i$-th row of $H$.
	
	The decoder estimates the erased bits $\widehat{\mathbf{c}}^{\mathcal{E}}$ from the unerased bits $\widehat{\mathbf{c}}^{\mathcal{V}} = \mathbf{c}^{\mathcal{V}}$. Thus, \eqref{eq:BEC_decoder_LE_H_1} can be represented by the following linear equations:
	\begin{equation}\label{eq:BEC_decoder_LE_H}
		\left(H^{\mathcal{E}} \right)^T \widehat{\mathbf{c}}^{\mathcal{E}} = \mathbf{q}
	\end{equation}
	where $\mathbf{q} = \left(H^{\mathcal{V}} \right)^T \mathbf{c}^{\mathcal{V}}$ and $\left(H^{\mathcal{E}}\right)^T \in \mathbb{F}_2^{(n - k) \times e}$.

	\begin{remark}\label{BEC:overdetermined}
		In \eqref{eq:BEC_decoder_LE} and \eqref{eq:BEC_decoder_LE_H}, the number of equations is more than or equal to the number of unknowns. Usually, these systems of linear equations are \emph{overdetermined}. The reason is that $k \le n-e$ for correcting $e$ erasures. Note that $G^{\mathcal{V}} \in \mathbb{F}_2^{(n-e) \times k}$ and $\left(H^{\mathcal{E}}\right)^T \in \mathbb{F}_2^{(n - k) \times e}$. Note that \eqref{eq:BEC_decoder_LE} and \eqref{eq:BEC_decoder_LE_H} are \emph{consistent} linear systems (i.e., there is at least one solution). 
	\end{remark}	
	
	%Since $\dim \left( \mathcal{C} \right) = k$, there exists exactly one solution of \eqref{eq:BEC_decoder_LE} so long as $\rank \left(G^{\mathcal{V}}\right) = k$. If $\rank \left(G^{\mathcal{V}}\right) < k$, there are several solutions, which may result in decoding failure. Similarly, there exists exactly one solution of \eqref{eq:BEC_decoder_LE_H} so long as $\rank \left( H^{\mathcal{E}}\right) = e$. Otherwise, there are several solutions, which may result in decoding failure.
	
	The minimum distance $d$ of $\mathcal{C}$ is given by
	\begin{align} \label{eq:BEC_dmin}
		d &= \underset{
			\substack{
				\mathbf{x} \ne \mathbf{0} \\
				H^T \mathbf{x}= \mathbf{0}
			}}
			{\text{min }} \|\mathbf{x}\|
	\end{align}
	which shows that any $d - 1$ rows of $H$ are linearly independent. So \eqref{eq:BEC_decoder_LE_H} has a unique solution when e is less than $d$.  
		
	The following Lemma has been known in coding theory community.	
	\begin{lemma}\label{lemma:BEC_UB} The upper bound on the probability of decoding failure of the MAP decoding rule is given by
		\begin{equation}\label{eq:BEC_UB}
			P\left(D=0 \mid |\mathcal{E}|=e \right) \le \frac{\sum_{w=d}^{e}{A_w \binom{n-w}{e-w}}}{\binom{n}{e}}
		\end{equation}
		where $A_w$ is the weight distribution of $\mathcal{C}$. 
	\end{lemma}
	\begin{IEEEproof}
		The proof was well known, which can be found in~\cite{Kim2014duality}. 		
	\end{IEEEproof}
%	\begin{IEEEproof}
%		Without loss of generality, we can assume that the all-zero codeword $\mathbf{0}$ has been transmitted and there exists a nonzero codeword $\mathbf{c}$ of Hamming weight $w$ such that $\Psi_w(\mathbf{c}) \subseteq \mathcal{E}$ where $\Psi_w(\mathbf{c})=\left\{ i \mid c_i \ne 0 \right\}$ denotes the locations of nonzero elements of $\mathbf{c}$ and $\mathcal{E}=\left\{i_1, \cdots, i_e \right\}$ denotes the locations of $e$ erasures. From the given decoding rule, $\mathbf{y}$ agrees with two codewords $\mathbf{0}$ and $\mathbf{c}$ on unerased bits, which may result in decoding failure. Meanwhile, if there is no nonzero codeword $\mathbf{c}$ such that $\Psi_w(\mathbf{c}) \subseteq \mathcal{E}$, then $\mathbf{y}$ agrees with only $\mathbf{0}$ on the unerased bits and the decoding succeeds.
%		
%		For a nonzero $\mathbf{c}$ such that $\Psi_w(\mathbf{c}) \subseteq \mathcal{E}$, the number of possible $\mathcal{E}$ is $\binom{n-w}{e-w}$. Due to double counting, the number of possible $\mathcal{E}$ which results in decoding failure will be less than or equal to $\sum_{w=d}^{e}{A_w \binom{n-w}{e-w}}$. Since the number of all possible $\mathcal{E}$ such that $|\mathcal{E}|=e$ is $\binom{n}{e}$, the upper bound on $P\left(D=0 \mid |\mathcal{E}|=e \right)$ is given by \eqref{eq:BEC_UB}.
%	\end{IEEEproof}
	
	$P\left(D=0 \mid |\mathcal{E}|=e \right)$ can be obtained exactly for $d \le e \le d + \left\lfloor \frac{d-1}{2} \right\rfloor$ (where $\left\lfloor x \right\rfloor$ represents the largest integer not greater than $x$) as stated in the following Lemma.
	
	\begin{lemma}\cite{Kim2014duality}\label{lemma:BEC_exact} For $e \le d + t$ where $t = \left\lfloor \frac{d-1}{2} \right\rfloor$, we can show that
		\begin{equation} \label{eq:BEC_exact}
			P\left(D=0 \mid |\mathcal{E}|=e \right) = \frac{1}{2} \cdot \frac{\sum_{w=d}^{e}{A_w \binom{n-w}{e-w}}}{\binom{n}{e}}.
		\end{equation}
	\end{lemma}
%	\begin{IEEEproof}
%		The proof can be found in~\cite{Kim2014duality}. 		
%	\end{IEEEproof}
	
	From the definition of $d$ in \eqref{eq:BEC_dmin}, Lemma~\ref{lemma:BEC_UB} and Lemma~\ref{lemma:BEC_exact}, we can state the following. 	
	\begin{theorem}\cite{Kim2014duality}\label{thm:BEC_bound} $P\left(D = 0 \mid |\mathcal{E}|=e\right)$ is given by
	\begin{numcases}{}
		0 &  for $e < d$,
		\\
		\frac{1}{2} \cdot \frac{\sum_{w=d}^{e}{A_{w} \binom{n-w}{e-w}}}{\binom{n}{e}} &  for $d \le e \le d+t$,
		\\
		\le \frac{\sum_{w=d}^{e}{A_{w} \binom{n-w}{e-w}}}{\binom{n}{e}} &  for $ e > d+t$.
	\end{numcases}
	\end{theorem}
			
	\subsection{Binary Defect Channel}
	
	We now summarize the defect channel model~\cite{Kuznetsov1974}. Define a variable $\lambda$ that indicates whether the memory cell is defective or not and $\widetilde{\mathbb{F}}_2 = \mathbb{F}_2 \cup \{\lambda \}$. Let ``$\circ$'' denote the operator  $\circ:\mathbb{F}_2 \times \widetilde{\mathbb{F}}_2 \rightarrow \mathbb{F}_2$ as in~\cite{Heegard1983plbc}
	\begin{equation}\label{eq:circ_operator}
	x \circ s =
	\begin{cases}
	x, & \text{if } s = \lambda ; \\
	s, & \text{if } s \ne \lambda.
	\end{cases}
	\end{equation}
	By using the operator $\circ$, an $n$-cell memory with defects is modeled by
	\begin{equation}
	\mathbf{y} = \mathbf{x} \circ \mathbf{s} \label{eq:BDC_vector}
	\end{equation}
	where $\mathbf{x}, \mathbf{y}\in \mathbb{F}_2^n$ are the channel input and output vectors. Also, the channel state vector $\mathbf{s} \in \widetilde{\mathbb{F}}_2^n$ represents the defect information in the $n$-cell memory. Note that $\circ$ is the vector component-wise operator. 
	
	If $s_i = \lambda$, this $i$-th cell is called \emph{normal}. If the $i$-th cell is \emph{defective} (i.e., $s_i \ne \lambda$), its output $y_i$ is stuck-at $s_i$ independent of the input $x_i$. So, the $i$-th cell is called stuck-at defect whose stuck-at value is $s_i$. The probabilities of stuck-at defects and normal cells are given by
	\begin{equation}\label{eq:defect_probability}
	P(S = s) =
	\begin{cases}
	1 - \beta, & \text{if } s = \lambda ; \\
	\frac{\beta}{2}, & \text{if } s = 0~\text{or}~1
	\end{cases}
	\end{equation}
	where the probability of stuck-at defects is $\beta$. Fig.~\ref{fig:BDC} shows the binary defect channel for $q=2$. 

	The number of defects is equal to the number of non-$\lambda$ components in $\mathbf{s}$. The number of errors due to defects is given by
	\begin{equation}\label{eq:BDC_num_errors}
		\| \mathbf{x} \circ \mathbf{s} - \mathbf{x} \|.
	\end{equation}
		
	The goal of masking stuck-at defects is to make a codeword whose values at the locations of defects match the stuck-at values of corresponding  defects \cite{Kuznetsov1974, Tsybakov1975additive}. The additive encoding and its decoding can be formulated as follows.
	
	\emph{Encoding:} A message $\mathbf{m} \in \mathbb{F}_2^k$ is encoded to a corresponding codeword $\mathbf{c}$ by
	\begin{equation}\label{eq:additive_encoding}
	\mathbf{c} = (\mathbf{m}, \mathbf{0}_{n-k}) + \mathbf{c}_0 = (\mathbf{m}, \mathbf{0}_{n-k}) + G_0 \mathbf{p}
	\end{equation}
	where $G_0 \in \mathbb{F}_2^{n \times (n-k)}$. By adding $\mathbf{c}_0 = G_0 \mathbf{p} \in \mathcal{C}_0$, we can mask defects among $n$ cells. Since the channel state vector $\mathbf{s}$ is available at the encoder, the encoder should choose $\mathbf{p} \in \mathbb{F}_2^{n-k}$ judiciously. The optimal parity $\mathbf{p}$ is chosen to minimize the number of errors due to defects, i.e., $ \|\mathbf{c} \circ \mathbf{s} - \mathbf{c} \|$.
	
	\emph{Decoding:} The decoding can be given by
	\begin{equation} \label{eq:decoding}
	\widehat{\mathbf{m}} = H_0^T \mathbf{y}
	\end{equation}
	where $\widehat{\mathbf{m}}$ represents the recovered message of $\mathbf{m}$. Note that the parity check matrix $H_0$ of $\mathcal{C}_0$ is given by $H_0 = [I_k \quad R]^T$ and $H_0^T G_0 = \mathbf{0}_{k, n-k}$. Note that \eqref{eq:decoding} is equivalent to the equation of coset codes. 
	
%	For the systematic codes, $G_0$ is given by~\cite{Heegard1983plbc}
%	\begin{equation}
%	G_0 = \begin{bmatrix} R \\ I_{n-k} \end{bmatrix}
%	\end{equation}
%	where $R \in \mathbb{F}_2^{k \times (n-k)}$ and $I_{n-k}$ is the $(n-k)$-dimensional identity matrix. Note that the identity matrix is located in the parity part unlike the conventional error-control codes. 
	
	The encoder knows the channel state vector $\mathbf{s}$ and tries to minimize $ \|\mathbf{c} \circ \mathbf{s} - \mathbf{c} \|$ by choosing $\mathbf{p}$ judiciously. Heegard proposed the minimum distance encoding (MDE) as follows \cite{Heegard1983plbc}.
		\begin{align}
			\mathbf{p}^*  & =  \underset{\mathbf{p}}{\text{argmin }} \left\| \mathbf{c}^{\mathcal{U}} - \mathbf{s}^{\mathcal{U}} \right\| \nonumber \\
			&= \underset{\mathbf{p}}{\text{argmin }} \left\| G_0^{\mathcal{U}} \mathbf{p} + \mathbf{b}^{\mathcal{U}} \right\| \label{eq:BDC_opt_problem}
	\end{align}
	where $\mathcal{U}=\left\{i_1,\cdots,i_u \right\}$ indicates the set of locations of $u$ defects. Also, $\mathbf{c}^{\mathcal{U}}=\left(c_{i_1},\cdots,c_{i_u}\right)$, $\mathbf{s}^{\mathcal{U}}=\left(s_{i_1},\cdots,s_{i_u}\right)$, and $G_0^{\mathcal{U}}=\left[\mathbf{g}_{0,i_1}^T,\cdots,\mathbf{g}_{0,i_u}^T \right]^T$. Since $\mathbf{b} =  (\mathbf{m}, \mathbf{0}_{n-k}) - \mathbf{s}$, $\mathbf{b}^{\mathcal{U}}$ is given by
	\begin{equation}\label{eq:BDC_b}
		\mathbf{b}^{\mathcal{U}} =  (\mathbf{m}, \mathbf{0}_{n-k})^{\mathcal{U}} - \mathbf{s}^{\mathcal{U}}.
	\end{equation}
	Note that $\left\|G_0^{\mathcal{U}} \mathbf{p}+ \mathbf{b}^{\mathcal{U}} \right\|$ represents the number of errors due to defects which is equal to the number in \eqref{eq:BDC_num_errors}.
	
	By solving the optimization problem of \eqref{eq:BDC_opt_problem}, the number of errors due to defects will be minimized. Also, Heegard showed that the MDE achieves the capacity~\cite{Heegard1983plbc}. However, the computational complexity for solving \eqref{eq:BDC_opt_problem} is exponential, which is impractical. Hence, we consider a polynomial time encoding approach. Instead of the MDE, we just try to solve the following linear equation~\cite{Tsybakov1975additive}.
	\begin{equation}\label{eq:BDC_encoder_LE}
		G_0^{\mathcal{U}} \mathbf{p} = \mathbf{b}^{\mathcal{U}}
	\end{equation}
	where $G_0^{\mathcal{U}} \in \mathbb{F}_2^{u \times (n-k)}$. Gaussian elimination or some other linear equation solution methods can be used to solve \eqref{eq:BDC_encoder_LE} with $\mathcal{O}\left(n^3\right)$ (i.e., $\mathcal{O}\left(n^3\right)$ due to $u \simeq \beta n$). If the encoder fails to find a solution of \eqref{eq:BDC_encoder_LE}, then an encoding failure is declared.
	
	For convenience, we define a random variable $E$ as follows.
	\begin{equation} \label{eq:BDC_E}
	E =
	\begin{cases}
	1, & \|\mathbf{c} \circ \mathbf{s} - \mathbf{c} \| = 0 \text{ (encoding success)} \\
	0, & \|\mathbf{c} \circ \mathbf{s} - \mathbf{c} \| \ne 0 \text{ (encoding failure)}
	%		1, & \eqref{eq:BDC_encoder_LE}\text{ has a solution (encoding success)} \\
	%		0, & \text{ otherwise (encoding failure)}
	\end{cases}
	\end{equation}	
	
	We can see that the probability of encoding failure $P(E=0)$ by the MDE of \eqref{eq:BDC_opt_problem} is the same as $P(E=0)$ by solving \eqref{eq:BDC_encoder_LE}. It is because $G_0^{\mathcal{U}} \mathbf{d} \ne  \mathbf{b}^{\mathcal{U}}$ if and only if $\|\mathbf{c} \circ \mathbf{s} - \mathbf{c} \| \ne 0$. Thus, $C_{\text{BDC}}$ can be achieved by solving \eqref{eq:BDC_encoder_LE}, which is easily shown by using the results of~\cite{Heegard1983plbc,Dumer1990}. 
	
	The coset coding of binning scheme can be described as solving the following linear equations~\cite{Wyner1974recent,Zamir2002}.
	\begin{equation} \label{eq:BDC_random_binning_1}
		H_0^T \mathbf{c} = \mathbf{m}
	\end{equation}
	where $\mathbf{c}$ is chosen to satisfy $\mathbf{c} \circ \mathbf{s} = \mathbf{c}$. \eqref{eq:BDC_random_binning_1} can be modified into
	\begin{align}\label{eq:BDC_random_binning_2}
	H_0^T \mathbf{c} & = \left(H_0^{\mathcal{U}} \right)^T \mathbf{c}^{\mathcal{U}} + \left(H_0^{\mathcal{W}} \right)^T \mathbf{c}^{\mathcal{W}} = \mathbf{m}
	\end{align}
	where $\mathcal{W} = \left\{j_1,\cdots, j_w\right\}$ represents the locations of normal cells such that $\mathcal{U} \cup \mathcal{W} = [n]$ and $\mathcal{U} \cap \mathcal{W} = \emptyset$. Note that $\mathbf{c}^{\mathcal{U}}=\left(c_{i_1}, \cdots, c_{i_u}\right)^T$, $\mathbf{c}^{\mathcal{W}}=\left(c_{j_1}, \cdots, c_{j_w}\right)^T$, $H_0^{\mathcal{U}}=\left[ \mathbf{h}_{0, i_1}^T, \cdots, \mathbf{h}_{0, i_u}^T \right]^T$ and $H_0^{\mathcal{W}}=\left[ \mathbf{h}_{0, j_1}^T, \cdots, \mathbf{h}_{0, j_w}^T \right]^T$ where $\mathbf{h}_{0, i}$ is the $i$-th row of $H_0$. Since $\mathbf{s}^{\mathcal{U}}$ is known to the encoder, the encoder can set $\mathbf{c}^{\mathcal{U}} = \mathbf{s}^{\mathcal{U}}$. Thus, the coset coding can be described as solving the following linear equation.
	\begin{equation}\label{eq:BDC_encoder_LE_binning}
	\left(H_0^{\mathcal{W}} \right)^T \mathbf{c}^{\mathcal{W}} = \mathbf{m}'
	\end{equation}
	where $\mathbf{m}' = \mathbf{m} - \left(H_0^{\mathcal{U}} \right)^T \mathbf{s}^{\mathcal{U}}$. The solution of \eqref{eq:BDC_encoder_LE_binning} represents the codeword elements of normal cells. Note that $\left(H_0^{\mathcal{W}} \right)^T \in \mathbb{F}_2^{k \times (n - u)}$. 
	
	\begin{remark}\label{BDC:underdetermined}
		In \eqref{eq:BDC_encoder_LE} and \eqref{eq:BDC_encoder_LE_binning}, the number of equations is less than or equal to the number of unknowns. Usually, these systems of linear equations are \emph{underdetermined}.
		The reason is that $k \le n-u$ for masking $u$ defects~\cite{Kuznetsov1974}. Note that $G_0^{\mathcal{U}} \in \mathbb{F}_2^{u \times (n-k)}$ and $\left(H_0^{\mathcal{W}} \right)^T \in \mathbb{F}_2^{k \times (n - u)}$. If \eqref{eq:BDC_encoder_LE} and \eqref{eq:BDC_encoder_LE_binning} have more than one solution, we can mask $u$ defects by choosing one of them. We can see the duality between Remark~\ref{BEC:overdetermined} and Remark~\ref{BDC:underdetermined}. 
	\end{remark}		

	The \emph{minimum distance} of additive encoding is given by 
	\begin{align} \label{eq:BDC_dmin}
	d^{\star} &= \underset{
		\substack{
			\mathbf{x} \ne \mathbf{0} \\
			G_0^T \mathbf{x}= \mathbf{0}
		}}
		{\text{min }} \|\mathbf{x}\|
	\end{align}
	which means that any $d^{\star} - 1$ rows of $G_0$ are linearly independent. Thus, additive encoding guarantees masking up to $d^{\star}-1$ stuck-at defects~\cite{Tsybakov1975additive, Heegard1983plbc}. 

	Similar to Lemma~\ref{lemma:BEC_UB}, we can derive the upper bound on the probability of encoding failure for $u$ defects.		
	\begin{lemma}\cite{Kim2013coding}\label{lemma:BDC_UB} The upper bound on $P(E=0||\mathcal{U}|=u)$ is given by
		\begin{equation}\label{eq:BDC_UB}
			P\left(E=0 \mid |\mathcal{U}|=u \right) \le \frac{\sum_{w=d^{\star}}^{u}{B_{w} \binom{n-w}{u-w}}}{\binom{n}{u}}
		\end{equation}
		where $B_{w}$ is the weight distribution of $\mathcal{C}_{0}^{\perp}$ (i.e., the dual code of $\mathcal{C}_0$).
	\end{lemma}

	%It is worth mentioning that the upper bound on $P(E=0 \mid |\mathcal{U}|=u)$ for the BDC is similar to the upper bound on $P\left(D=0 \mid |\mathcal{E}|=e\right)$ for the BEC presented in Lemma~\ref{lemma:BEC_UB}.
		
	The following Lemma states that $P\left(E=0 \mid |\mathcal{U}|=u \right)$ can be obtained exactly for $d^{\star} \le u \le d^{\star} + \left\lfloor \frac{d^{\star}-1}{2} \right\rfloor$.
	
	\begin{lemma}\cite{Kim2013coding}\label{lemma:BDC_exact} For $u \le d^{\star} + t^{\star}$ where $t^{\star} = \left\lfloor \frac{d^{\star} - 1}{2} \right\rfloor$, $P\left(E=0 \mid |\mathcal{U}|=u \right)$ is given by
		\begin{equation}
			\label{eq:BDC_exact_condition}
			P\left(E=0 \mid |\mathcal{U}|=u \right) = \frac{1}{2} \cdot \frac{\sum_{w=d^{\star}}^{u}{B_{w} \binom{n-w}{u-w}}}{\binom{n}{u}}.
		\end{equation}
	\end{lemma}

	Similar to the upper bound on $P\left( D = 0 \mid |\mathcal{E}|=e \right)$ in Theorem~\ref{thm:BEC_bound} for the BEC, we can provide the upper bound on $P\left( E=0 \mid |\mathcal{U}|=u \right)$ for the BDC as follows.
	
	\begin{theorem}\cite{Kim2013coding} \label{thm:BDC_bound} $P\left(E=0 \mid |\mathcal{U}|=u\right)$ is given by
		\begin{numcases}{}
			0 & for $u < d^{\star}$,
			\\
			\frac{1}{2} \cdot \frac{\sum_{w=d^{\star}}^{u}{B_{w} \binom{n-w}{u-w}}}{\binom{n}{u}}  & for $d^{\star} \le u \le d^{\star}+t^{\star}$, \label{eq:BDC_exact_condition_thm}
			\\
			\le \frac{\sum_{w=d^{\star}}^{u}{B_{w} \binom{n-w}{u-w}}}{\binom{n}{u}}  & for $ u > d^{\star}+t^{\star}$.
		\end{numcases}
	\end{theorem}
	
	By comparing Theorem~\ref{thm:BEC_bound} and Theorem~\ref{thm:BDC_bound}, the duality of erasures and defects can be seen. We will discuss this duality in the following subsection. 

	\subsection{Duality between Erasures and Defects}
	
	\begin{table*}[t!]
		\renewcommand{\arraystretch}{1.3}
		\caption{Duality between BEC and BDC}
		\label{tab:duality}
		\centering
		{\small
			\hfill{}
			\begin{tabular}{|c|c|c|}
				\hline
				& BEC & BDC  \\ \hline \hline
				%Channel              & $P(\text{erasure}) = \alpha$ & $P(\text{defect}) = \beta$        \\ \hline
				\multirow{2}{*}{Channel property} & Ternary output $Y \in \{0, 1, * \}$ & Ternary state $S \in \{0, 1, \lambda\}$  \\ 
				& (erasure $*$ is neither ``0'' nor ``1'') & (defect is either ``0'' or ``1'') \\ \hline
				Capacity             & $C_{\mathrm{BEC}} = 1 - \alpha$ \: \eqref{eq:BEC_capacity} & $C_{\mathrm{BDC}} = 1 - \beta$ \: \eqref{eq:BDC_capacity} \\ \hline
				Channel state information  & Locations & Locations and stuck-at values \\ \hline
				Correcting / Masking & Decoder corrects erasures & Encoder masks defects    \\ \hline
				\multirow{3}{*}{MAP decoding / MDE} &  $G^{\mathcal{V}} \widehat{\mathbf{m}} =\mathbf{y}^{\mathcal{V}}$\: \eqref{eq:BEC_decoder_LE} & $G_0^{\mathcal{U}} \mathbf{p} = \mathbf{b}^{\mathcal{U}}$ \:\eqref{eq:BDC_encoder_LE} \\
				& $\left(H^{\mathcal{E}} \right)^T \widehat{\mathbf{c}}^{\mathcal{E}} = \mathbf{q}$ \ \eqref{eq:BEC_decoder_LE_H} & $ \left(H_0^{\mathcal{W}} \right)^T \mathbf{c}^{\mathcal{W}} = \mathbf{m}'$ \ \eqref{eq:BDC_encoder_LE_binning}   \\
				& (Overdetermined) & (Underdetermined) \ \\ \hline
				\multirow{2}{*}{Solutions} & $\widehat{\mathbf{m}}$ (estimate of message) or & $\mathbf{p}$ (parity) or \\
				& $\widehat{\mathbf{c}}^{\mathcal{E}}$ (estimate of erased bits) & $\mathbf{c}^{\mathcal{W}}$ (codeword elements of normal cells) \\ \hline
				\multirow{2}{*}{Minimum distance} & $d = \min\{ \|\mathbf{x}\|: H^T \mathbf{x} = \mathbf{0}, \mathbf{x} \ne \mathbf{0} \}$ & $d^{\star} = \min\{ \|\mathbf{x}\|: G_0^T \mathbf{x} = \mathbf{0}, \mathbf{x} \ne \mathbf{0} \}$ \\
				& If $e < d$, $e$ erasures are corrected. & If $u < d^{\star}$, $u$ defects are masked. \\ \hline
				Upper bounds on& \multirow{2}{*}{Theorem~\ref{thm:BEC_bound}}  & \multirow{2}{*}{Theorem~\ref{thm:BDC_bound}} \\
				probability of failure & & \\ \hline
				Probability of failure & \multicolumn{2}{c|}{If $H = G_0$ and $\alpha = \beta$, then $P(D = 0) = P(E = 0)$ (Theorem~\ref{thm:duality}) } \\ \hline		
			\end{tabular}}
		\hfill{}
%		\vspace{-3mm}
	\end{table*}	
		
	We will discuss the duality of erasures and defects as summarized in Table~\ref{tab:duality}. In the BEC, the channel input $X \in \left\{0, 1 \right\}$ is binary and the channel output $Y = \left\{ 0, 1, * \right\}$ is ternary where the erasure $*$ is \emph{neither} 0 nor 1. In the BDC, the channel state $S \in \left\{0, 1, \lambda \right\}$ is ternary whereas the channel input and output are binary. The ternary channel state $S$ informs whether the given cells are stuck-at defects or normal cells. The stuck-at value is \emph{either} 0 or 1.
	
	The expressions for capacities of both channels are quite similar as shown in \eqref{eq:BEC_capacity} and \eqref{eq:BDC_capacity}. In the BEC, the \emph{decoder} corrects erasures by using the information of locations of erasures, whereas the \emph{encoder} masks the defects by using the information of defect locations and stuck-at values in the BDC.
	
	The capacity achieving scheme of the BEC can be represented by the linear equations based on the generator matrix $G$ of \eqref{eq:BEC_decoder_LE} or the linear equations based on the parity check matrix $H$ of \eqref{eq:BEC_decoder_LE_H}. Both linear equations are usually \emph{overdetermined} as discussed in Remark~\ref{BEC:overdetermined}. The solution of linear equations based on $G$ is the \emph{estimate of message} $\widehat{\mathbf{m}}$ and there should be only one $\widehat{\mathbf{m}}$ for decoding success. Also, the solution of linear equations based on $H$ is the \emph{estimate of erased bits} $\widehat{\mathbf{c}}^{\mathcal{E}}$ which should be only one $\widehat{\mathbf{c}}^{\mathcal{E}}$ for decoding success.
	
	On the other hand, the capacity achieving scheme of the BDC can be described by the linear equation which are usually \emph{underdetermined} as explained in~\ref{BDC:underdetermined}. The additive encoding can be represented by the linear equations based on the generator matrix $G_0$ of \eqref{eq:BDC_encoder_LE} whose solution is the \emph{parity} $\mathbf{p}$. Also, the binning scheme can be represented by the linear equations based on the parity check matrix $H_0$ of \eqref{eq:BDC_encoder_LE_binning} whose solution is the \emph{codeword elements of normal cells} $\mathbf{c}^{\mathcal{W}}$. Unlike the coding scheme of the BEC, there can be several solutions of $\mathbf{p}$ or $\mathbf{c}^{\mathcal{W}}$ that mask all stuck-at defects.
	
	We can see the duality between erasures and defects by comparing the solution $\widehat{\mathbf{m}}$ of \eqref{eq:BEC_decoder_LE} and the solution $\mathbf{p}$ of \eqref{eq:BDC_encoder_LE}, i.e., \emph{message} and \emph{parity}. Note that coding schemes of \eqref{eq:BEC_decoder_LE} and \eqref{eq:BDC_encoder_LE} are based on the generator matrix. In addition, we can compare the duality of \emph{codeword elements of erasures} and \emph{codeword elements of normal cells} from \eqref{eq:BDC_encoder_LE} and \eqref{eq:BDC_encoder_LE_binning} which are coding schemes based on the parity check matrix.
	
	In the BEC, the minimum distance $d$ is defined by the \emph{parity check matrix} $H$, whereas the minimum distance $d^{\star}$ of the BDC is defined by the \emph{generator matrix} $G_0$. The upper bound on the probability of decoding failure is dependent on the weight distribution of $\mathcal{C}$ (i.e., $A_w$), whereas the upper bound on the probability of encoding failure is dependent on the weight distribution of $\mathcal{C}_0^{\perp}$ (i.e., $B_w$).
	
	If $A_w = B_w$ and $e = u$, it is clear that the upper bound on $P\left( D=0 \mid |\mathcal{E}|=e \right)$ is same as the upper bound on $P\left( E=0 \mid |\mathcal{U}|=u \right)$ by Theorem~\ref{thm:BEC_bound} and Theorem~\ref{thm:BDC_bound}. In particular, the following Theorem shows the equivalence of the failure probabilities (i.e., the probability of decoding failure of erasures and the probability of encoding failure of defects).
	
	\begin{theorem}\cite{Kim2014duality}\label{thm:duality} If $H = G_0$ and $\alpha = \beta$, then the probability of decoding failure of MAP decoding for the BEC is the same as the probability of encoding failure of MDE for the BDC (i.e., $P(D=0) = P(E=0)$). The complexity for both is $\mathcal{O}(n^3)$.
	\end{theorem}
	\begin{IEEEproof}
		If $\alpha = \beta$, then it is clear that that $P(\mathcal{E}) = P(\mathcal{U})$ for $\mathcal{E} = \mathcal{U}$. If $\mathcal{E} = \mathcal{U}$ and $H = G_0$, then $H^{\mathcal{E}} = G_0^{\mathcal{U}}$. If $H^{\mathcal{E}}$ and $G_0^{\mathcal{U}}$ are full rank, then it is clear that $P(D=0)=P(E=0)=0$.
		
		Suppose that $\rank(H^{\mathcal{E}}) = \rank(G_0^{\mathcal{U}}) = e - j$ where $\mathcal{E} = \mathcal{U}$ (i.e., $e = u$). For the BEC, there are $2^j$ codewords that satisfy \eqref{eq:BEC_decoder_LE_H} and the decoder chooses one codeword among them randomly. Hence, $P(D=0 \mid \mathcal{E})=1 - \frac{1}{2^j}$. 
		
		For the BDC, each element of $\mathbf{b}^{\mathcal{U}}$ in \eqref{eq:BDC_encoder_LE} is uniform since $P(S=0 \mid S \ne \lambda) = P(S=1 \mid S \ne \lambda) = \frac{1}{2}$. \eqref{eq:BDC_encoder_LE} has at least one solution if and only if $\rank(G_0^{\mathcal{U}}) = \rank(G_0^{\mathcal{U}} \mid \mathbf{b}^{\mathcal{U}})$. In order to satisfy this condition, the last $j$ elements of $\mathbf{b}^{\mathcal{U}}$ should be zeros, which means that $P(E=0 \mid \mathcal{U})=1 - \frac{1}{2^j}$. Thus, $P(D=0 \mid \mathcal{E}) = P(E=0 \mid \mathcal{U})$ if $\mathcal{E} = \mathcal{U}$ and $H = G_0$. 
		
		Since $P(\mathcal{E}) = P(\mathcal{U})$ and $P(D=0 \mid \mathcal{E}) = P(E=0 \mid \mathcal{U})$ for $\mathcal{E}=\mathcal{U}$, it is true that $P(D=0) = P(E=0)$. 
	\end{IEEEproof}
	
\begin{figure}[!t]
		\centering
		\includegraphics[width=0.40\textwidth]{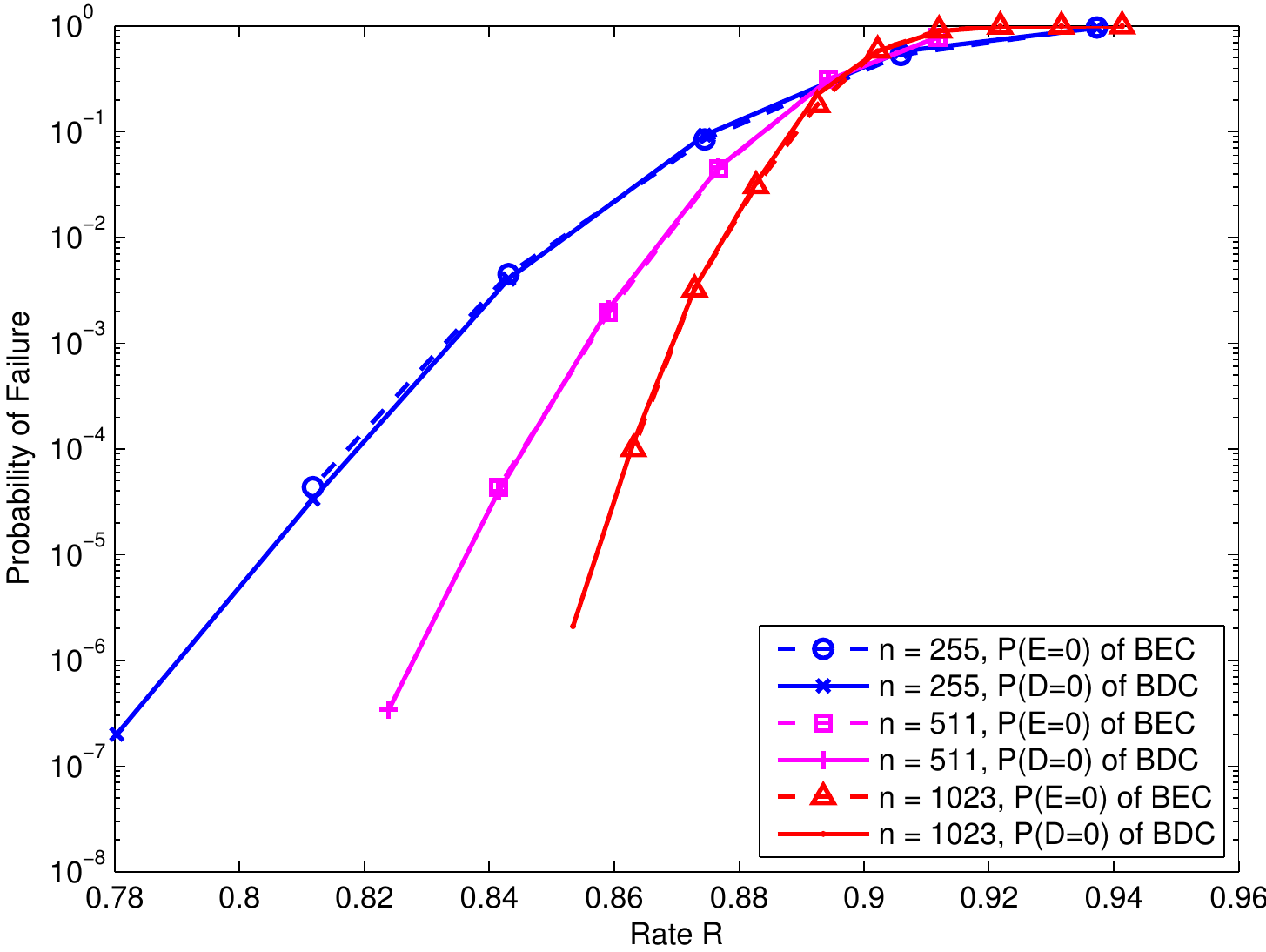}
		\caption{Probability of failure, i.e., $P(D=0)$ of the BEC with $\alpha=0.1$ and $P(E=0)$ of the BDC with $\beta=0.1$.}
		\label{fig:BEC_BDC_numerical}
		\vspace{-5mm}
	\end{figure}

	Fig.~\ref{fig:BEC_BDC_numerical} compares $P(D=0)$ of the BEC and $P(E=0)$ of the BDC when $H = G_0$ and $\alpha = \beta$. The parity check matrices of Bose-Chaudhuri-Hocquenghem (BCH) codes are used for $H$ and $G_0$. Hence, BCH codes are used for the BEC and the duals of BCH codes are used for the BDC. The numerical results in Fig.~\ref{fig:BEC_BDC_numerical} shows that $P(D=0) = P(E=0)$ if $H = G_0$ and $\alpha = \beta$, which confirms Theorem~\ref{thm:duality}. 
	
	Recently, it was proved that a sequence of linear codes achieves $C_{\text{BEC}}$ under MAP decoding if its blocklengths are strictly increasing, its code rates converge to some $\delta\in(0,1)$, and the permutation group of each code is doubly transitive~\cite{Kudekar2016}. Hence, RM codes and BCH codes can achieve $C_{\text{BEC}}$ under MAP decoding. Based on the duality between the BEC and the BDC, we can claim the following Corollary.  
	
	\begin{corollary}\label{thm:duality_rm}
		RM codes achieve $C_{\text{BDC}}$ with computational complexity $\mathcal{O}(n^3)$. 
	\end{corollary}	
	\begin{IEEEproof}
		In~\cite{Kudekar2016}, it was shown that RM codes achieve $C_{\text{BEC}}$ under MAP decoding whose computational complexity is $\mathcal{O}(n^3)$. By Theorem~\ref{thm:duality}, the duals of RM codes achieve $C_{\text{BDC}}$ with $\mathcal{O}(n^3)$. Since the duals of RM codes are also RM codes~\cite[pp. 375--376]{Macwilliams1977theory}, RM codes achieve $C_{\text{BDC}}$. 
	\end{IEEEproof}				
	We note that the duals of BCH codes also achieve $C_{\text{BDC}}$ by the same reason. 
		
	In this section, we have demonstrated the duality between the BEC and the BDC from channel properties, capacities, capacity-achieving schemes, and their failure probabilities. This duality implies that the existing code constructions and algorithms of the BEC can be applied to the BDC and \emph{vice versa}. 
	
\subsection{Relations between BEC, BDC, BEQ, and WOM} \label{subsec:relations}
	
	We extend the duality of the BEC and the BDC to other interesting models such as binary erasure quantization (BEQ) and write-once memory (WOM) codes. We review the literature on these models and describe the relations between BEC, BDC, BEQ, and WOM.  	
	
	\begin{figure}[!t]
		\centering
		\includegraphics[width=0.45\textwidth]{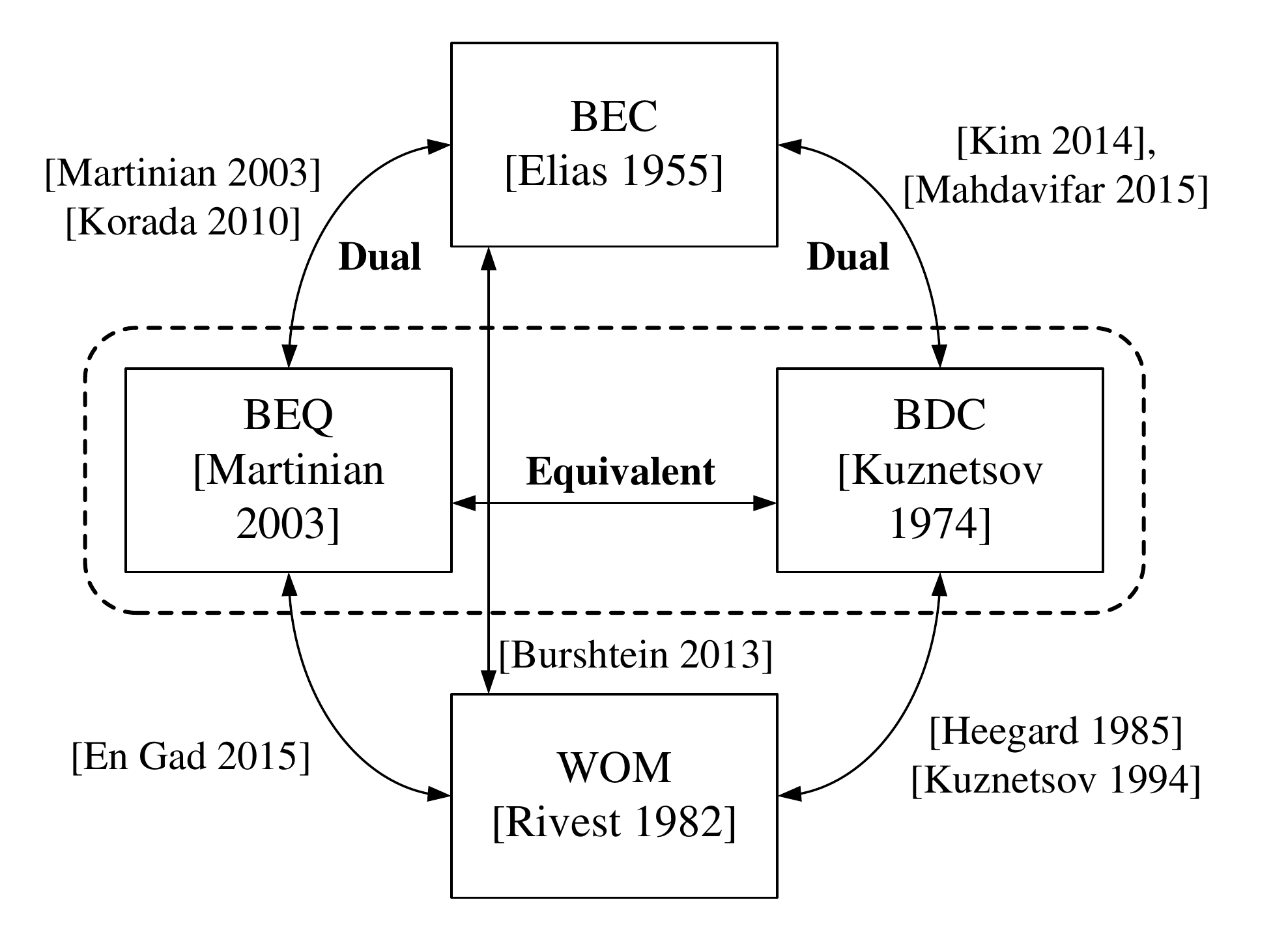}
		\caption{Relations between BEC, BDC, BEQ, and WOM.}
		\label{fig:connection}
		\vspace{-5mm}
	\end{figure}	
		
	Martinian and Yedidia~\cite{Martinian2003iterative} considered BEQ problems where the source vector consists of $\{0, 1, *\}$ ($*$ denotes an erasure). Neither ones nor zeros may be changed, but erasures may be quantized to either zero or one. The erasures do not affect the distortion regardless of the value they are assigned since erasures represents source samples which are missing, irrelevant, or corrupted by noise. The BEQ problem with erasure probability $\alpha$ can be formulated as follows. 
	\begin{equation}\label{eq:BEQ_probability}
	P(S = s) =
	\begin{cases}
	\alpha, & \text{if } s = * ; \\
	\frac{1 - \alpha}{2}, & \text{if } s= 0~\text{or}~1
	\end{cases}
	\end{equation}
	and the Hamming distortion $d_H(\cdot, \cdot)$ is given by
	\begin{equation}\label{eq:BEQ_distortion}
	d_H(0, *) = d_H(1, *) = 0, \quad d_H(0, 1)=1.
	\end{equation}
	
	The rate-distortion bound with zero distortion is given by
	\begin{equation}\label{eq:BEQ_bound}
	R_\text{BEQ} = 1 - \alpha.
	\end{equation}
	In~\cite{Martinian2003iterative} the duality between the BEC and the BEQ was observed, and the authors showed that low-density generator matrix (LDGM) codes (i.e., the duals of LDPC codes) can achieve the $R_\text{BEQ}$ by modified message-passing algorithm. The computational complexity is $\mathcal{O}(nd_{\mathcal{G}})$ where $d_{\mathcal{G}}$ denotes the maximum degree of the bipartite graph $\mathcal{G}$ of the low-density generator matrix. In~\cite{Korada2010source}, it was shown that polar codes with an successive cancellation encoder can achieve $R_\text{BEQ}$ with $\mathcal{O}(n \log n)$. 
	
	From \eqref{eq:BEQ_probability}--\eqref{eq:BEQ_bound}, we can claim that the BEQ with erasure probability $\alpha$ is equivalent to the BDC with defect probability $\beta$ if $\beta = 1 - \alpha$. We can observe that $s=*$ of the BEQ corresponds to $s = \lambda$ of the BDC, which represents normal cells by comparing \eqref{eq:defect_probability} and \eqref{eq:BEQ_probability}. Also, $s = 0$ and $s=1$ of the BEQ can be regarded as stuck-at 0 defects and stuck-at 1 defects, respectively. In addition, $R_\text{BEQ} = C_\text{BEC} = 1 - C_\text{BDC}$. 
	
	Since the BEQ is equivalent to the BDC, we can claim that RM codes achieve $R_\text{BEQ}$ due to Corollary~\ref{thm:duality_rm}. Hence, LDGM codes (duals of LDPC codes), polar codes, and RM codes achieve $R_{\text{BEQ}}$.  
	
	Inversely, the coding scheme for the BEQ can be applied to the BDC. Thus, $C_{\text{BDC}}$ can be achieved by LDGM codes and polar codes whose complexities are $\mathcal{O}(nd_{\mathcal{G}})$ and $\mathcal{O}(n \log n)$ respectively. It is important because the best known encoding complexity of capacity achieving scheme for the BDC was $\mathcal{O}(n \log^2 n)$ in~\cite{Dumer1990}. Also, note that the encoding complexity of coding schemes in~\cite{Mahdavifar2015} is $\mathcal{O}(n^3)$.   
		
	The model of WOM was proposed for data storage devices where once a one is written on a memory cell, this cell becomes permanently associated with a one. Hence, the ability to rewrite information in these memory cells is constrained by the existence of previously written ones~\cite{Rivest1982wom, Heegard1985}. Recently, the WOM model has received renewed attention as a possible channel model for flash memories due to their asymmetry between write and erase operations~\cite{Jiang2007wom, Yaakobi2012codes}.
		
	In~\cite{Heegard1985}, it was noted that WOM are related to the BDC since the cells storing ones can be considered as stuck-at 1 defects. Moreover, Kuznetsov and Han Vinck~\cite{Kuznetsov1994} showed that additive encoding for the BDC can be used to achieve the capacity of WOM. Burshtein and Strugatski~\cite{Burshtein2013polar} proposed a capacity-achieving coding scheme for WOM with $\mathcal{O}(n \log n)$ complexity, which is based on polar codes and successive cancellation encoding~\cite{Korada2010source}. Recently, En Gad \emph{et al.}~\cite{EnGad2015rewriting} related the WOM to the BEQ. Hence, LDGM codes and message-passing algorithm in~\cite{Martinian2003iterative} can be used for WOM. Note that the encoding complexity is $\mathcal{O}(nd_{\mathcal{G}})$.   
	
	Fig.~\ref{fig:connection} illustrates the relations between BEC, BDC, BEQ, and WOM. We emphasize that a coding scheme for one model can be applied to other models based on these relations. It is worth mentioning that RM codes, LDPC (or LDGM) codes, and polar codes can achieve the capacities of all these models. Their computational complexities are $\mathcal{O}(n^3)$, $\mathcal{O}(nd_{\mathcal{G}})$, and $\mathcal{O}(n \log n)$, respectively.

\section{Locally Rewritable Codes (LWC)}\label{sec:LWC}
 
	Inspired by the duality between erasures and defects, we proposed locally rewritable codes (LWC). LWC were introduced to improve endurance and power consumption of resistive memories which can be modeled by the BDC. After briefly reviewing resistive memories and LRC, we explain LWC and their properties. The details of LWC can be found in~\cite{Kim2016lwc}.  
 
\subsection{Resistive Memories}\label{subsec:resistive}

	Resistive memory technologies are promising since they are expected to offer higher density than dynamic random-access memories (DRAM) and better speed performance than NAND flash memories~\cite{xpoint2015}. Phase change memories (PCM) and resistive random-access memories (RRAM) are two major types of resistive memories. Both have attracted significant research interest due to their scalability, compactness, and simplicity. 
	
	The main challenges that prevent their large-scale deployment are endurance limit and power consumption~\cite{Wong2010PCM, Wong2012RRAM}. The endurance limit refers to the maximum number of writes before the memory becomes unreliable. Beyond the given endurance limit, resistive memory cells are likely to become defects~\cite{Kim2012PCM, Sharma2014RRAM}. In addition, the power consumption depends on the number of writes. Hence, the number of writes is the key parameter for reliability of memory cells and power efficiency. 

\subsection{Locally Repairable Codes (LRC)}\label{subsec:LRC}

	An $(n, k, d, r)$ LRC is a code of length $n$ with information (message) length $k$, minimum distance $d$, and repair locality $r$. If a symbol in the LRC-coded data is lost due to a node failure, its value can be repaired (i.e. reconstructed) by accessing at most $r$ other symbols~\cite{Gopalan2012, Tamo2014LRC}. 
	
	One way to ensure fast repair is to use low repair locality such that $r \ll k$ at the cost of minimum distance $d$. The relation between $d$ and $r$ is given by~\cite{Gopalan2012}
	\begin{equation} \label{eq:general_singleton}
		d \le n - k - \left\lceil \frac{k}{r} \right\rceil + 2.
	\end{equation}
	It is worth mentioning that this bound is a generalization of the Singleton bound. The LRC achieving this bound with equality are called \emph{optimal}. Constructions of the optimal LRC were proposed in~\cite{Silberstein2013, Tamo2014LRC, Tamo2013matroid}. %Recently, several binary LRC constructions have been proposed~\cite{Goparaju2014, Shahabinejad2014, Huang2015, Tamo2015subcodes, Silberstein2015}. 
	
\subsection{Locally Rewritable Codes}\label{subsec:LWC}

	As a toy example, suppose that $n$-cell \emph{binary} memory has a single stuck-at defect. It is easy to see that this stuck-at defect can be handled by the following simple technique~\cite{Kuznetsov1974}. 
	\begin{equation} \label{eq:single_defect}
	\mathbf{c} = ( \mathbf{m}, 0) + \mathbf{1}_n \cdot p
	\end{equation}
	where $G_0 = \mathbf{1}_n$. 
	
	Suppose that $i$-th cell is a defect whose stuck-at value is $s_i \in \mathbb{F}_2$. If $i \in [n-1]$ and $s_i = m_i$, or if $i=n$ and $s_n = 0$, then $p$ should be 0. Otherwise, $p=1$. 
	
	If there is no stuck-at defect among $n$ cells, then we can store $\mathbf{m}$ by writing $\mathbf{c} = (\mathbf{m}, 0)$ (i.e., $p=0$). Now, consider the case when stored information needs to be updated causing $\mathbf{m}$ to become $\mathbf{m}'$. Usually, $\| \mathbf{m}- \mathbf{m}' \| \ll n$, which happens often due to the updates of files. Instead of storing $\mathbf{m}'$ into another group of $n$ cells, it is more efficient to store $\mathbf{m}'$ by rewriting only  $\| \mathbf{m}- \mathbf{m}' \|$ cells. For example, suppose that $m_i' \ne m_i$ for an $i \in [k]$ and $m_j' = m_j$ for all other $j \in [k] \setminus i$. Then, we can store $k$-bit $\mathbf{m}'$ by rewriting only one cell. 
	
	An interesting problem arises when a cell to be rewritten is defective. Suppose that $i$-th cell is a stuck-at defect whose stuck-at value is $s_i$. If $s_i = m_i \ne m_i'$, then we should write $\mathbf{c} = (\mathbf{m}, 0)$ for storing $\mathbf{m}$. However, in order to store the updated information $\mathbf{m}'$, we should write $\mathbf{c}' = \overline{\mathbf{c}} = (\overline{\mathbf{m}}, 1)$ where $p=1$. Thus, $n-1$ cells should be rewritten to update one bit data $m_i'$ without stuck-at error. The same thing happens when $s_i = m_i' \ne m_i $. When considering endurance limit and power consumption, rewriting $n-1$ cells is a high price to pay for preventing one bit stuck-at error.  
	
	In order to relieve this burden, we can change \eqref{eq:single_defect} by introducing an additional parity bit as follows. 
	\begin{align} \label{eq:single_defect_locality}
	\mathbf{c} & = \left( \mathbf{m}_{[1:\frac{n}{2}]}, 0,\mathbf{m}_{[\frac{n}{2}+1:n]} , 0 \right) + G_0 \mathbf{p} \\
	&= \left( \mathbf{m}_{[1:\frac{n}{2}]}, 0,\mathbf{m}_{[\frac{n}{2}+1:n]} , 0 \right) + \begin{bmatrix} \mathbf{1}_{\frac{n}{2}} & \mathbf{0}_{\frac{n}{2}} \\ \mathbf{0}_{\frac{n}{2}} & \mathbf{1}_{\frac{n}{2}} \end{bmatrix} (p_1, p_2) 
	\end{align}
	where $k = n-2$. For simplicity's sake, we assume that $n$ is even. Then, $\mathbf{1}_{\frac{n}{2}}$ and $\mathbf{0}_{\frac{n}{2}}$ are all-ones and all-zeros column vectors with ${n}/{2}$ elements. By introducing an additional parity bit, we can reduce the number of rewriting cells from $n-1$ to $\frac{n}{2}-1$. 
	
	This idea is similar to the concept of Pyramid codes which are the early LRC~\cite{Huang2007pyramid}. For $n$ disk nodes, single parity check codes can repair one node failure (i.e., single erasure) by 
	\begin{equation} \label{eq:single_erasure}
	\mathbf{1}_n^T \widehat{\mathbf{c}} = 0
	\end{equation} 
	where $\widehat{\mathbf{c}}$ represents the recovered codeword from disk node failures. Assuming that $c_i$ is erased due to a node failure, $c_i$ can be recovered by 
	\begin{equation} \label{eq:single_erasure_repair}
	\widehat{c}_i = c_i = \sum_{j \in [n]\setminus i}{c_j}.
	\end{equation}
	For this recovery, we should access $k = n-1$ nodes which degrades the repair speed. For more efficient repair process, we can add a new parity as follows. 
	\begin{equation} \label{eq:single_erasure_locality}
	H^T \widehat{\mathbf{c}} = \begin{bmatrix} \mathbf{1}_{\frac{n}{2}} & \mathbf{0}_{\frac{n}{2}} \\ \mathbf{0}_{\frac{n}{2}} & \mathbf{1}_{\frac{n}{2}} \end{bmatrix}^T \widehat{\mathbf{c}} = \mathbf{0}
	\end{equation} 
	Then, a failed node $c_i$ can be repaired by accessing only $\frac{n}{2} - 1$ nodes. Note that the repair locality of \eqref{eq:single_erasure_locality} is $\frac{n}{2}-1$ whereas the repair locality of \eqref{eq:single_erasure} is $n - 1$ which is a simple but effecitve idea of Pyramid codes.   
	
	An interesting observation is that $G_0$ of \eqref{eq:single_defect_locality} is the same as $H$ of \eqref{eq:single_erasure_locality}. In addition, note that the number of resistive memory cells to be rewritten is the same as the number of nodes to be accessed in distributed storage systems. These observations can be connected to the duality between erasures and defects in Section~\ref{sec:duality}.
	
	We define \emph{initial writing cost} and \emph{rewriting cost} which are related to write endurance and power consumption. 
	
	\begin{definition} [Initial Writing Cost] Suppose that $\mathbf{m}$ was stored by its codeword $\mathbf{c}$ in the initial stage of $n$ cells where all the normal cells are set to zeros. The writing cost is given by
		\begin{equation} \label{eq:writing_cost_def}
			\Delta (\mathbf{m}) = \| \mathbf{c}  \| - u_{\setminus 0}
		\end{equation}
		where $u_{\setminus 0}$ denotes the number of stuck-at defects whose stuck-at values are nonzero. 
	\end{definition}
	
	In \eqref{eq:writing_cost_def}, we assume that there are $u$ defects among $n$ cells and $\mathbf{c}$ masks these $u$ stuck-at defects successfully. So, we do not need to write stuck-at defects since their stuck-at values are the same as corresponding elements of $\mathbf{c}$. 
	
	\begin{definition} [Rewriting Cost] Suppose that $\mathbf{m}$ was stored by its codeword $\mathbf{c}$ in $n$ cells. If $\mathbf{c}'$ is rewritten to these $n$ cells to store the updated $\mathbf{m}'$, the rewriting cost is given by
		\begin{equation} \label{eq:rewriting_cost_def}
			\Delta (\mathbf{m}, \mathbf{m}') = \| \mathbf{c} - \mathbf{c}' \| 
		\end{equation}
		where we assume that both $\mathbf{c}$ and $\mathbf{c}'$ mask stuck-at defects. 
	\end{definition}
	
	High rewriting cost implies that the states of lots of cells should be changed, which is harmful to endurance and power efficiency. 
	
	%It is worth mentioning that, in general, the rewriting cost is more important than the initial writing cost since most of write operations will be rewriting. If a device offers write endurance of 10000 cycles, the write operations of 9999 will be rewriting whereas only one among 10000 writing is the initial write operation (i.e., 0.01\%). However, there may be some storage applications (such as for archival storage), where the number of initial writings and rewritings may be similar.
	
	Now, we introduce the \emph{rewriting locality} which affects initial writing cost and rewriting cost. The rewriting locality is a counterpart of repair locality of LRC. As repair locality is meaningful for a single disk failure, rewriting locality is valid when there is a single stuck-at defect among $n$ cells. In distributed storage systems, the most common case is a single node failure among $n$ nodes~\cite{Huang2007pyramid}. Similarly, for a proper defect probability $\beta$, we can claim that the most common scenario of resistive memories is that there is a single defect among $n$ cells.

	\begin{definition} [Information Rewriting Locality] \label{def:inf_r} Suppose that $m_i$ for $i \in [k]$, i.e., information (message) part, should be updated to $m_i' \ne m_i$ and the corresponding $i$-th cell is a stuck-at defect. If $m_i$ can be updated to $m_i'$ by rewriting $r^{\star}$ other cells, then the $i$-th coordinate has \emph{information rewriting locality} $r^{\star}$. 
	\end{definition}
	
	\begin{lemma}\cite{Kim2016lwc}\label{thm:inf_r} If the $i$-th coordinate for $i \in [k]$ has information rewriting locality $r^{\star}$, then there exists $\mathbf{c}_0 \in \mathcal{C}_0$ such that $i \in \text{supp}(\mathbf{c}_0)$ and $\left\| \mathbf{c}_0 \right\| = r^{\star}+1$.
	\end{lemma}
%	\begin{IEEEproof}
%		For $\mathbf{m}$ and $\mathbf{m}'$, suppose that $m_i \ne m_i'$ for an $i \in [k]$ and $m_j = m_j'$ for all other $j \in [k] \setminus i$. Note that $i$-th cell is a stuck-at defect whose stuck-at value is $s_i$. We should consider the following cases:
%		\begin{enumerate}
%			\item 	$m_i' \ne m_i = s_i$. 
%			\item 	$m_i \ne m_i' = s_i$. 
%			\item 	$m_i \ne m_i'$, $m_i \ne s_i$ and $m_i' \ne s_i$. 
%		\end{enumerate}	
%		
%		For	$m_i' \ne m_i = s_i$, it is obvious that $\mathbf{c} = (\mathbf{m}, \mathbf{0}_{n-k})$ and $\mathbf{c}' = (\mathbf{m}', \mathbf{0}_{n-k})  + \mathbf{c}_0'$ where $c_i' = m_i' + c_{0, i}' = s_i$ by~\eqref{eq:additive_encoding}. For the information rewriting locality $r^{\star}$, $\mathbf{c}_0' \in \mathcal{C}_0$ should satisfy $\left\| \mathbf{c}_0' \right\| = r^{\star}+1$ to mask the stuck-at defect by writing $r^{\star}$ cells. Note that we do not need to write the stuck-at defect since its stuck-at value is $s_i = c_i'$. For $m_i \ne m_i' = s_i$, the proof is similar. 
%		
%		For $m_i \ne m_i'$, $m_i \ne s_i$ and $m_i' \ne s_i$, $\mathbf{c} = (\mathbf{m}, \mathbf{0}_{n-k}) + \mathbf{c}_0$ and $\mathbf{c}' = (\mathbf{m}', \mathbf{0}_{n-k})  + \mathbf{c}_0'$. We can pick $\mathbf{c}_0$ and $\mathbf{c}_0'$ such that $\mathbf{c}_0' = \alpha \mathbf{c}_0$ (where $\alpha \in \mathbb{F}_q $) and $m_i + c_{0, i} = m_i' + c_{0, i}' = s_i$. For the information rewriting locality $r^{\star}$, $\mathbf{c}_0$ and $\mathbf{c}_0'$ should satisfy $\left\| \mathbf{c}_0 \right\| = \left\| \mathbf{c}_0' \right\| = r^{\star} + 1$. 	
%	\end{IEEEproof}
	
	If a stuck-at defect's coordinate is $i \in [k+1:n]$, i.e. parity location, then $\mathbf{m}$ can be updated to $\mathbf{m}'$ by rewriting $\| \mathbf{m} - \mathbf{m}' \|$ cells because of $\mathbf{c}_0 = \mathbf{c}_0'$. Thus, a stuck-at defect in the parity location is not related to rewriting. However, a stuck-at defect in the parity location affects initial writing. We will define parity rewriting locality as follows.   
	
	\begin{definition} [Parity Rewriting Locality] \label{def:par_r} Suppose that only one nonzero symbol $m_i$ should be stored to the initial stage of $n$ cells. Note that there is a stuck-at defect in the parity location $j$ for $j \in [k+1:n]$ (i.e., parity part) and $s_j \ne 0$. If $m_i$ can be stored by writing at most $r^{\star} + 1$ cells, then the $j$-th coordinate has \emph{parity rewriting locality} $r^{\star}$. 
	\end{definition}
	
	\begin{lemma}\cite{Kim2016lwc}\label{thm:par_r} If the $j$-th coordinate for $j \in [k+1:n]$ has parity rewriting locality $r^{\star}$, then there exists $\mathbf{c}_0 \in \mathcal{C}_0$ such that $j \in \text{supp} (\mathbf{c}_0)$ and $\left\| \mathbf{c}_0 \right\| = r^{\star}+1$.
	\end{lemma}
%	\begin{IEEEproof}
%		Suppose that $\mathbf{m}$ should be stored to the initial stage of $n$ cells where $m_i \ne 0$ for $i \in [k]$ and $m_i' = 0$ for $i' \in [k] \setminus i$. By~\eqref{eq:additive_encoding}, $\mathbf{c} = (\mathbf{m}, \mathbf{0}_{n-k}) + \mathbf{c}_0 $ such that $c_{0, j} = s_j$ where the $j$-th cell is a stuck-at defect for $j \in [k+1:n]$. For the parity rewriting locality $r^{\star}$, $\mathbf{c}_0$ should satisfy $\left\| \mathbf{c}_0 \right\| = r^{\star}+1$. If $i \in \text{supp}(\mathbf{c}_0)$, then it is possible to store $m_i$ without stuck-at error by writing $\|\mathbf{c}_{0 \setminus j} \| = r^\star$ cells since we do not need to write $c_{0, j}$. Otherwise, we should write both $m_i$ and $\|\mathbf{c}_{0 \setminus j} \|$, i.e., $r^\star + 1$ cells. 	
%	\end{IEEEproof}
		
	\begin{definition} [Locally Rewritable Codes] If any $i$-th coordinate for $i \in [n]$ has (information or parity) rewriting locality at most $r^{\star}$, then this code is called locally rewritable code (LWC) with rewriting locality $r^{\star}$. $(n, k, d^{\star}, r^{\star})$ LWC code is a code of length $n$ with information length $k$, minimum distance $d^{\star}$, and rewriting locality $r^{\star}$.
	\end{definition}
	
	Now, we show in the following theorem that rewriting locality $r^{\star}$ is an important parameter for rewriting cost. 
	
	\begin{theorem} \label{thm:rewriting_cost} Suppose that $\mathbf{m}$ is updated to $\mathbf{m}'$ by LWC with rewriting locality $r^{\star}$. If there is a single stuck-at defect in $n$ cells, then the rewriting cost $\Delta (\mathbf{m}, \mathbf{m}')$ is given by
		\begin{equation} \label{eq:rewriting_cost_bound}
			\Delta (\mathbf{m}, \mathbf{m}') \le \|\mathbf{m} - \mathbf{m}' \| + r^{\star} - 1.
		\end{equation}
	\end{theorem}

	\begin{corollary}\cite{Kim2016lwc} \label{thm:writing_cost}If $\mathbf{m}$ is stored in the initial stage of $n$ cells with a single stuck-at defect, then the writing cost $\Delta (\mathbf{m})$ is given by
		\begin{equation} \label{eq:writing_cost}
			\Delta (\mathbf{m}) \le \|\mathbf{m} \| + r^{\star}.
		\end{equation}
	\end{corollary}
%	\begin{IEEEproof}
%		First suppose that the single defect's coordinate is $i \in [k]$ and its stuck-at value is $s_i$. If $s_i = m_i$, then $\mathbf{c} = (\mathbf{m}, \mathbf{0})$. The writing cost $\Delta (\mathbf{m}) = \| \mathbf{c} \| - 1 = \| \mathbf{m} \| - 1$. 
%		
%		If $s_i \ne m_i$, then $\mathbf{c} = (\mathbf{m}, \mathbf{0}) + \mathbf{c}_0$. Then, 
%		\begin{align} 
%			\Delta (\mathbf{m}) &= \|\mathbf{c} \| - 1 \\
%			& \le  \|\mathbf{m} \| + \| \mathbf{c}_0 \| - 1 \label{eq:writing_cost_pf0}\\
%			& =  \|\mathbf{m} \| + r^{\star} \label{eq:writing_cost_pf1}
%		\end{align}
%		where \eqref{eq:writing_cost_pf1} follows from Lemma~\ref{thm:inf_r}. 
%		
%		Next suppose that the single defect's coordinate is $j \in [k+1: n]$. If $s_j = 0$, then $\Delta (\mathbf{m}) = \|\mathbf{c}\| =\|(\mathbf{m}, \mathbf{0}) \| =  \|\mathbf{m}\|$. If $s_j \ne 0$, then $\mathbf{c} = (\mathbf{m}, \mathbf{0}) + \mathbf{c}_0$ where $j \in \text{supp}(\mathbf{c}_0)$. By Lemma~\ref{thm:par_r}, we can claim that $\Delta (\mathbf{m}) \le \|\mathbf{m} \| + \| \mathbf{c}_0 \| - 1 = \|\mathbf{m} \| + r^{\star} $. 
%	\end{IEEEproof}
	
	%In resistive memories, we can claim that rewriting cost is more important than initial writing cost because there is no erase operation as in flash memories. All the initial stage of cells would be used up after a while, hence most of data will be stored by rewriting resistive memory cells. 
	
	Theorem~\ref{thm:rewriting_cost} and Corollary~\ref{thm:writing_cost} show that a small rewriting locality $r^{*}$ can reduce writing cost and rewriting cost, which is helpful for improving endurance and power consumption. 
	
\subsection{Duality of LRC and LWC} \label{subsec:duality}
	
	In this subsection, we investigate the duality of LRC and LWC, which comes from the duality between erasures and defects in Section~\ref{sec:duality}. We show that existing construction methods of LRC can be used to construct LWC based on this duality. First, the relation between minimum distance $d^{\star}$ and rewriting locality $r^{\star}$ is observed. 
	
	%\begin{claim} \label{thm:dist_r}Let $\mathcal{C}_0$ denote a linear code whose generator matrix is $G_0$ of \eqref{eq:additive_encoding}. If there exists $\mathbf{c}_0 \in \mathcal{C}_0$ such that $i \in \text{supp}\left(\mathbf{c}_0 \right)$ and $\| \mathbf{c}_0 \| = r^{\star} + 1$, then the $i$-th element has rewriting locality $r^{\star}$.   
	%\end{claim}
	%\begin{IEEEproof}
	%While proving Theorem~\ref{thm:rewriting_cost} and Corollary~\ref{thm:writing_cost}, we use the condition of $\| \mathbf{c}_0 \| - 1 = r^{\star}$ where the $i$-th coordinate belongs to the support of $\mathbf{c}_0$. 
	%\end{IEEEproof}	
	
	\begin{definition} \label{def:cyc_LWC}If $\mathcal{C}_0$ is cyclic, then the LWC is called \emph{cyclic}. 
	\end{definition}
	
	\begin{lemma}\cite{Kim2016lwc}\label{thm:cyc} Let $\mathcal{C}_0$ denote a cyclic code whose minimum distance is $d_0$. Then, corresponding cyclic LWC's rewriting locality is $r^{\star} = d_0 - 1$.	
	\end{lemma}
%	\begin{IEEEproof}
%		Due to the property of cyclic codes, we can claim that there exists $\mathbf{c}_0 \in \mathcal{C}_0$ such that $i \in \text{supp}(\mathbf{c}_0)$ and $\| \mathbf{c}_0 \| = d_0$ for any $i \in [n]$. Since $d_0$ is the minimum distance of $\mathcal{C}_0$, we can claim that the rewriting locality is $r^{\star} = d_0 - 1$. 
%	\end{IEEEproof}
	
	From the definition of $d^{\star}$ in \eqref{eq:BDC_dmin}, $d^{\star} = d_0^{\perp}$ which is the minimum distance of $\mathcal{C}_0^{\perp}$, namely, dual code of $\mathcal{C}_0$. Thus, the parameters of cyclic LWC is given by
	\begin{equation} \label{eq:LWC_parameter}
		(d^{\star}, r^{\star}) = (d_0^{\perp}, d_0 - 1). 
	\end{equation}
	In \cite{Huang2015, Tamo2015subcodes}, an equivalent relation for cyclic LRC was given by
	\begin{equation} \label{eq:LRC_parameter}
		(d, r) = (d, d^{\perp}-1).
	\end{equation}
	
	By comparing \eqref{eq:LWC_parameter} and \eqref{eq:LRC_parameter}, we observed the \emph{duality} between LRC and LWC. This duality is important since it indicates that we can construct LWC using existing construction methods of LRC as shown in the following theorem. 
	
	\begin{theorem}\cite{Kim2016lwc}\label{thm:LWC_construction}
		Suppose that $H_{\text{LRC}} \in \mathbb{F}_q^{n \times (n-k)}$ is the parity check matrix of cyclic LRC $\mathcal{C}_{\text{LRC}}$ with $(d, r) = (d, d^{\perp} - 1)$. By setting $G_0 = H_{\text{LRC}}$, we can construct cyclic LWC $\mathcal{C}_{\text{LWC}}$ with 
		\begin{equation}
			(d^{\star}, r^{\star}) = (d, d^{\perp} - 1).
		\end{equation}
	\end{theorem}
%	\begin{IEEEproof}
%		By setting $G_0 = H_{\text{LRC}}$, the LWC's codeword $\mathbf{c} \in \mathcal{C}_{\text{LWC}}$ is given by
%		\begin{equation*}\label{eq:additive_encoding_dual}
%			\mathbf{c} = (\mathbf{m}, \mathbf{0}) + H_{\text{LRC}} \cdot \mathbf{p}. 
%		\end{equation*}
%		The minimum distance $d^{\star}$ of $\mathcal{C}_{\text{LWC}}$ is given by
%		\begin{align*} \label{eq:dmin_dual}
%			d^{\star} &= \underset{
%				\substack{
%					\mathbf{x} \ne \mathbf{0} \\
%					H_{\text{LRC}}^T \mathbf{x}= \mathbf{0}
%				}}
%				{\text{min }} \|\mathbf{x}\|
%		\end{align*}
%		which is equivalent to the minimum distance $d$ of $\mathcal{C}_{\text{LRC}}$. Hence, we can claim that
%		\begin{equation} \label{eq:dmin_relation}
%			d^{\star}=d_0^{\perp} = d.
%		\end{equation} 
%		From \eqref{eq:LWC_parameter} and \eqref{eq:dmin_relation}, $r^{\star}=d_0 - 1 = d^{\perp} - 1$.
%	\end{IEEEproof}
	
	In Theorem~\ref{thm:duality}, we showed that the decoding failure probability of the optimal decoding scheme for the BEC is the same as the encoding failure probability of the optimal encoding scheme for the BDC. By setting $H = G_0$, capacity-achieving codes for the BDC can be constructed from state of art codes for the BEC. Similarly, Theorem~\ref{thm:LWC_construction} shows that we can construct $(n, k, d^{\star}=d, r^{\star}=d^{\perp}-1)$ LWC by using existing construction methods of $(n, k, d, r=d^{\perp}-1)$ LRC.

	\begin{remark}[Optimal Cyclic LWC]\cite{Kim2016lwc} \label{thm:LWC_optimal}
		Theorem~\ref{thm:LWC_construction} shows that the optimal cyclic $(n, k, r, d)$ LRC can be used to construct the optimal cylic $(n, k, r^\star, d^\star)$ LWC such that
		\begin{equation}
			d^{\star} = n - k - \left \lceil \frac{k}{r^{\star}} \right \rceil + 2.
		\end{equation} 
	\end{remark}
	Hence, the optimal LWC can be constructed from the optimal LRC. 
	
	\begin{remark}[Bound of LWC] \cite{Kim2016lwc}
		From Theorem~\ref{thm:LWC_construction} and Remark~\ref{thm:LWC_optimal}, we can claim the following bound for LWC. 
		\begin{equation}
			d^{\star} \le n - k - \left \lceil \frac{k}{r^{\star}} \right \rceil + 2
		\end{equation}
		which is equivalent to the bound for LRC given by \eqref{eq:general_singleton}. 		
	\end{remark}
	
	\begin{table}[t]
		\renewcommand{\arraystretch}{1.4}
		\caption{Duality of LRC and LWC}
		\label{tab:duality_LRC_LWC}
		\centering
		{\hfill{}
			\begin{tabular}{|c|c|c|}
				\hline
				& $(n, k, d, r)$ LRC    & $(n, k, d^{\star}, r^{\star})$ LWC \\ \hline \hline
				\multirow{2}{*}{Application}     & Distributed storage systems & Resistive memories \\ 
				& (system level)              & (physical level) \\ \hline
				Channel   & Erasure channel             & Defect channel   \\ \hline
				Encoding        & $\mathbf{c} = G_{\text{LRC}}\mathbf{m}$ & $\mathbf{c} = (\mathbf{m}, \mathbf{0})+ H_{\text{LRC}}\mathbf{p}$   \\ \hline
				Decoding        & $H_{\text{LRC}}^T \widehat{\mathbf{c}} = \mathbf{0}$ & $G_{\text{LRC}}^T \mathbf{c} = \widehat{\mathbf{m}}$  \\ \hline
				Bound           & $d \le n - k - \left \lceil \frac{k}{r} \right \rceil + 2$ & $d^{\star} \le n - k - \left \lceil \frac{k}{r^{\star}} \right \rceil + 2$ \\ \hline
				\multirow{2}{*}{Trade-off} & $d$ (reliability) vs. & $d^{\star}$ (reliability) vs. \\
				& $r$ (repair efficiency) & $r^*$ (rewriting cost)  \\ \hline
			\end{tabular}}
			\hfill{}
			\vspace{-3mm}
	\end{table}
		
	In Table~\ref{tab:duality_LRC_LWC}, the duality properties of LRC and LWC are summarized, which comes from the duality of the BEC and the BDC.

%\begin{figure}[!t]
%\centering
%\subfloat[Single-level cell (SLC) for $B=1$]{\includegraphics[width=3in]{fig_slc.eps}
%\label{fig:slc}}
%\hfil
%\vspace{-1mm}
%\subfloat[Multi-level cell (MLC) for $B=2$]{\includegraphics[width=3in]{fig_mlc.eps}
%\label{fig:mlc}}
%\caption{Threshold voltage distribution of flash memory cells.}
%\label{fig:slc_mlc}
%\vspace{-5mm}
%\end{figure}

%\begin{figure}[t]
%   \centering
%   \includegraphics[width=0.29\textwidth]{fig_defect_channel.eps}
%   \caption{Channel model of a memory with defective cells.}
%   \label{fig:defect_channel}
%   \vspace{-5mm}
%\end{figure}

\section{Conclusion}\label{sec:conclusion}

The duality between the BEC and the BDC was investigated. We showed that RM codes and duals of BCH codes achieve the capacity of the BDC based on this duality. This duality can be extended to the relations between BEC, BDC, BEQ, and WOM. 

Based on these relations, we showed that RM codes achieve the capacity of the BDC with $\mathcal{O}(n^3)$ and LDGM codes (duals of LDPC codes) achieve the capacity with $\mathcal{O}(nd_{\mathcal{G}})$. Also, polar codes can achieve the capacity with $\mathcal{O}(n \log n)$ complexity, which beats the best known result of $\mathcal{O}(n \log^2 n)$.   

Also, we proposed the LWC for resistive memories based on this duality, which are the counterparts of LRC for distributed storage systems. The proposed LWC can improve endurance limit and power consumption which are major challenges for resistive memories.

%% Appendix:
%% If needed a single appendix is created by
%\appendix
%% If several appendices are needed, then the command
%\appendices
%% in combination with further \section-commands can be used.

%\appendix

%%% Use \section* for acknowledgement
%\section*{Acknowledgment}
%
%The authors would like to thank various sponsors for supporting
%their research.

%% References:
%% We recommend the usage of BibTeX:
%%

\IEEEtriggeratref{38}

\bibliographystyle{IEEEtran}
\bibliography{IEEEabrv,duality_ita_bib}

\end{document}